\documentclass[twocolumn,showpacs,preprintnumbers,draftclsnofoot]{revtex4}
\usepackage{amsmath}
\usepackage{amssymb}
\usepackage{graphicx}
\usepackage{dcolumn}
\usepackage{bm}
\usepackage{amssymb}
\usepackage{graphicx}
\usepackage{dcolumn}
\usepackage{bm}
\begin{document}

\title{ Built-In Electric Field Modulation of Spontaneous Magnetism and Thermospin Transport in Boron-Nitrogen Doped Zigzag Graphene Nanoribbons  }
\author{Ma Luo\footnote{Corresponding author:luoma@gpnu.edu.cn}, Yiyi Xie and Xinxiu Wu }
\affiliation{School of Physics and Optoelectronic Engineering, Guangdong Polytechnic Normal University, Guangzhou 510665, China}

\begin{abstract}

Zigzag-edged graphene nanoribbons (ZGNRs) host large magnetic moments at two zigzag edges due to spontaneous magnetism, whose ground state is antiferromagnetically coupled spin-order edge state (AF state). In this paper, the spontaneous magnetism and thermoelectricity of ZGNRs with periodic substituted doping of boron and nitrogen atoms along the axial direction are investigated by first principle calculation. The doping induces an internal electric field, which modulates the magnetic moments at the two zigzag edges and the band structure with spin splitting. For varying doping configuration, the ground state could be AF state, ferromagnetically coupled spin-order edge state (FM state), or degenerated pair of AF and FM states. External transversal static electric field can further modulate the magnetic moments and the band structure. By designing the doping configuration and the external field, pure thermal spin current with the absence of thermal charge current can be generated at room temperature, which can be switched on and off by flipping the magnetic configuration between AF and FM states. Thus, the doped ZGNRs can be applied as prototype devices for spin-caloritronic.

\end{abstract}

\maketitle

\section{Introduction}

Graphene nanoribbons (GNRs) possess tunable bandgaps, ultrahigh carrier mobility, and robust edge states, rendering them promising building blocks for post-Moore microelectronics \cite{reviewGNR1,Zutic04}. Owing to these superior properties, GNRs have been extensively explored for the development of low-power-high-speed transistors \cite{reviewGNR2}, interconnecting circuits \cite{Areshkin10,YueeXie12}, spin-electronic devices \cite{WHan14}, opto-electronic devices \cite{Zamani18}, and spin-caloritronics devices \cite{seebeck001,seebeck002,seebeck003,seebeck004,seebeck005,seebeck006,seebeck007,seebeck008,seebeck009,seebeck010,seebeck011,seebeck012,seebeck013}. Recent advances in the atomic-precision synthesis and monolithic integration of GNRs have further unlocked their tremendous potential in nanoscale logic devices and quantum components, paving a solid way for the development of next-generation integrated circuits \cite{reviewGNR3,Kristians2020}.

Zigzag graphene nanoribbons (ZGNRs) feature localized edge states near the Fermi level, where unpaired $\pi$-electrons exhibit spontaneous spin polarization induced by strong on-site Coulomb interactions. In accordance with Lieb$^{\prime}$s theorem \cite{lieb01,lieb02,lieb03}, the ground state of semi-infinite graphene sheets with zigzag edges sustains prominent magnetic moments at zigzag terminations, while the magnetic moments at other lattice sites remain small. Furthermore, the magnetic moment of each lattice site is aligned antiparallel to those of its nearest neighbors. For finite-width ZGNRs with balanced lattice site numbers on the two sublattices, the zigzag terminations on the two edges belong to distinct sublattices. As a result, the system favors an antiferromagnetic (AF) ground state, wherein the magnetic moments at the two zigzag edges are antiparallelly oriented  \cite{Mitsutaka96,Hikihara03,Yamashiro03,YoungWooSon06,YoungWoo06,Pisani07,Wunsch08,FernandezRossier08,Jung09,Rhim09,Lakshmi09,Jung09a,Yazyev10,Hancock10,Jung10,Manuel10,Feldner11,DavidLuitz11,JeilJung11,Culchac11,Schmidt12,Soriano12,Karimi12,Schmidt13,Golor13,Bhowmick13,FengHuang13,Ilyasov13,Carvalho14,Lado14,MichaelGolor14,PrasadGoli16,Baldwin16,Ortiz16,Hagymasi16,Ozdemir16,Friedman17,ZhengShi17,XiaoLong18,Krompiewski17,Krompiewski19,maluo2020,maluo2021}. In contrast, the first metastable excited state corresponds to the ferromagnetic (FM) state, characterized by parallel-aligned edge magnetic moments. Structural modification of zigzag edge geometries can break the sublattice symmetry by inducing an imbalance in lattice site numbers between the two sublattices, enabling ZGNRs to accommodate either FM or non-magnetic ground states \cite{shapeTune1,shapeTune2,shapeTune3}. Additionally, the magnetic strength of ZGNRs can be effectively modulated via mechanical twisting \cite{mechanicalTune1}, strain engineering and adatom doping \cite{adatomTune}, as well as multiferroic coupling effects \cite{ferroelectricTune,ferroelectricTune1}.

The magnetic moments of ZGNRs are stabilized by intrinsic lattice symmetry and the weak spin-orbit coupling inherent to carbon materials, facilitating long-lived spin coherence \cite{spinGraphene1,spinGraphene2}. Recent advancements in experimental techniques have addressed the long-standing instability issue of zigzag edges \cite{zigunstable} through atomically precise edge functionalization \cite{edgefunctionalizationTheory,edgefunctionalization} or hexagonal boron nitride (hBN) encapsulation \cite{ZGNR2}, enabling the direct experimental observation of room-temperature edge magnetism. Given that the band structure and electronic transport properties of ZGNRs are predominantly determined by their magnetic configurations, numerous strategies have been proposed to leverage their intrinsic magnetism for the design of logic-storage units and functional spintronic devices, including spin valves \cite{spinvalve} and spintronic switches \cite{spinswitch,spinswitch1}. Benefiting from the inherent edge-localized magnetism and highly tunable spin-dependent thermoelectric responses, ZGNRs serve as a compelling platform for spin-caloritronic research, offering great potential for developing low-dissipation thermal-spin conversion devices that operate purely under temperature gradients \cite{applicationSeeback}. A variety of structural and electronic modulation approaches have been developed to tailor the spin splitting of GNRs and further enhance the spin Seebeck effect, such as introducing antidot architectures within GNR lattices \cite{seebeck003,seebeck004}, engineering edge geometric configurations \cite{seebeck005,seebeck007,seebeck008,seebeck009}, implementing mechanical twisting \cite{seebeck006}, and modulating the Fermi level via doping manipulation \cite{seebeck011}.

This work systematically investigates the magnetic modulation of ZGNRs via axially aligned substitutional doping with boron-nitrogen (B-N) atomic arrays. Such doping configuration induces a transverse internal electric field, and the thermally driven spin current transport in the doped ZGNRs is further explored in detail. The doped systems retain overall electrical neutrality due to the equal number of doped B and N atoms. The spatial distribution of dopants precisely governs the profile of the transverse internal electric field and the local potential distribution at the two zigzag edges. The modulation effect of this intrinsic transverse electric field is analogous to that of externally applied transverse electric fields, which induce opposite energy level shifts at the two zigzag terminations \cite{YoungWooSon06}. For ZGNRs with distinct doping configurations, the total energy difference between antiferromagnetic (AF) and ferromagnetic (FM) states can be positive, negative, or nearly zero. The application of an additional external transverse electric field can further reshape the internal electric field distribution, thereby effectively tuning the magnetic configuration of ZGNRs. Notably, ZGNRs with energetically degenerate AF and FM ground states possess identical total energy at equilibrium, enabling their application as high-performance nanoscale logic-in-memory devices. Under an applied temperature gradient, the external transverse electric field can modulate the band structure of ZGNRs, thereby regulating both thermal charge and spin transport behaviors. At a specific temperature, the thermal charge current is completely suppressed, and switching between the degenerate AF and FM states achieves over ten-fold modulation of the pure thermal spin current. Consequently, the B-N atomic array-doped ZGNRs proposed in this study emerge as promising candidates for high-efficiency thermally driven logic spintronic devices.

The remainder of this paper is organized as follows. Section II systematically analyzes the variations in total energy, magnetic configurations, and band structures of AF and FM states modulated by doping geometries and external electric fields. Section III presents the calculation and discussion of thermal charge and spin currents for ZGNRs with representative doping configurations. Finally, Section IV gives the conclusion.

\section{Magnetic configuration and band structure}

\begin{figure}[tbp]
\centering
\scalebox{0.58}{\includegraphics{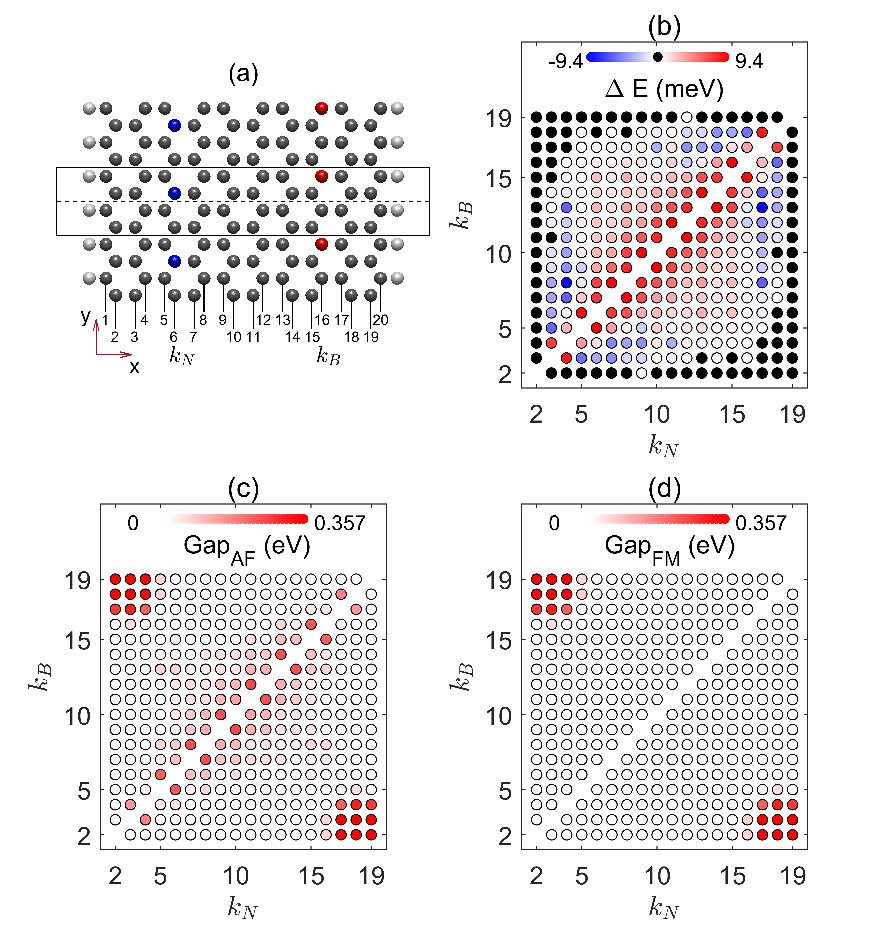}}
\caption{ (a) Structure of ZGNR with periodic substituted doping of nitrogen and boron atoms. One period of the doped ZGNR is surrounded by the square rectangular of solid line, which is separated by the dashed line into two unit cell of the pristine counterpart. The carbon, nitrogen, boron and hydrogen atoms are plotted as black, blue, red and write spheres. The indices of columns with substituted doping of nitrogen and boron atoms are marked as $k_{N}$ and $k_{B}$, respectively. (b) The difference between the total energy level of the FM and AF configuration versus $k_{N}$ and $k_{B}$ are indicated by the color. The band gap of the FM and AF configurations versus $k_{N}$ and $k_{B}$ are indicated by the color in (c) and (d), respectively. }
\label{figure_GapList}
\end{figure}

\begin{figure*}[tbp]
\centering
\scalebox{0.73}{\includegraphics{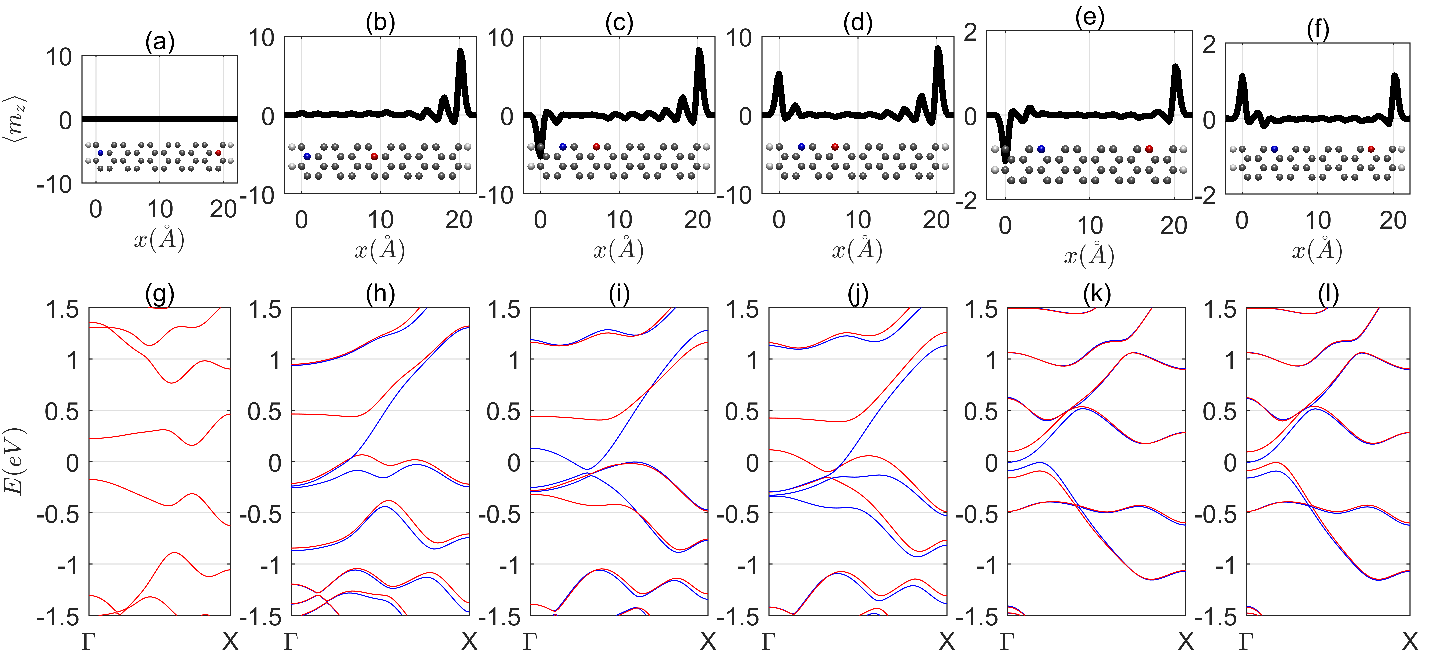}}
\caption{ In the first row, the spatial distribution of the spin polarization along x direction are plotted, with the inserted panel indicating the doping configuration. The corresponding band structures of spin up and down are plotted as blue and red lines, respectively, in the second row. The doping configuration $(k_{N},k_{B})$ are $(2,19)$ in (a,g), $(2,10)$ in (b,h), $(4,8)$ in (c,d,i,j), $(5,17)$ in (e,f,k,l).  }
\label{figure_BandList}
\end{figure*}

The structural geometry of axially aligned boron-nitrogen (B-N) array-doped ZGNRs is illustrated in Fig. \ref{figure_GapList}(a). In this work, ZGNRs with a width of 20 carbon atoms are adopted as the research model. For the doped system, B and N substitutional dopants are linearly arranged at a density of one dopant atom per two unit cells of pristine ZGNRs. Accordingly, the supercell of the doped ZGNR is constructed by extending two unit cells of the pristine ZGNR, as delineated by the dashed line in Fig. \ref{figure_GapList}(a). All numerical simulations in this study primarily focus on doped ZGNR configurations where B and N dopants are confined within a single pristine ZGNR unit cell. Configurations with B and N dopants distributed across different unit cells yield qualitatively consistent numerical results, which are presented in the appendix for reference. The doping geometry is defined by the coordinate pair ($k_{N}$,$k_{B}$), where $k_{N}$ and $k_{B}$ denote the column indices of N and B dopant atoms, respectively. To avoid overlapping atomic sites, the condition $k_{N}\ne k_{B}$ is strictly satisfied. The precise positions of N and B atoms within one supercell are further specified by the position vectors ($\mathbf{r}_{N}$,$\mathbf{r}_{B}$). This study concentrates on internal doping configurations, wherein neither B nor N atoms reside on the zigzag terminations, corresponding to the index range $1<k_{N}(k_{B})<20$. The electronic structures of pristine and doped ZGNRs are calculated via density functional theory (DFT) implemented in the quantumATK software package \cite{theDFT001,theDFT002}. The simulation supercell adopts a dimension of $(L_{x},L_{y},L_{z})=(36,4.922,30)$ $\AA$ along the $x$, $y$, and $z$ directions. The generalized gradient approximation parameterized by the Perdew-Burke-Ernzerhof (GGA-PBE) functional is employed to describe electronic exchange and correlation effects \cite{theDFT003,theDFT004}, combined with norm-conserving pseudopotentials \cite{theDFT005}. A plane-wave cutoff energy of 600 eV is set for all calculations, and the Brillouin zone is sampled using a $1\times15\times1$ Monkhorst-Pack k-point mesh. Geometric structural relaxation is performed until the residual atomic force converges below $0.01$ $eV/\AA$ and the total energy convergence threshold reaches $10^{-5}$ $eV$. For boundary condition settings, Neumann boundary conditions for the Poisson solver are applied to the top and bottom surfaces at $z=0$ and $L_{z}$ to mimic free-standing ZGNRs, while periodic boundary conditions are imposed on the $x$- and $y$-direction surfaces at $x=0$, $L_{x}$ and $y=0$, $L_{y}$. To obtain the electronic structures of antiferromagnetic (AF) and ferromagnetic (FM) states, the initial magnetic moments of carbon atoms at the two zigzag edges (sites 1 and 20) are set to antiparallel and parallel alignments, respectively. For systems under an external longitudinal electric field along the $x$-direction, two metallic electrode regions are defined within the ranges $x\in[0,0.36]$ $\AA$ and $x\in[35.64,36]$ $\AA$, spanning the full $y$ and $z$ dimensions. A finite voltage bias is applied between the two electrodes, and Dirichlet boundary conditions are adopted for the $x$-direction surfaces at $x=0$ and $L_{x}$ to characterize the field-induced electronic modulation.

\subsection{Modulation by doping configuration}

The total energies of the FM and AF states in a single unit cell, denoted as $E_{FM}$ and $E_{AF}$, are extracted from DFT calculations. In the absence of an external electric field, the energy difference $\Delta E\equiv E_{FM}-E_{AF}$ as a function of doping configuration $(k_{N},k_{B})$ is visualized via the color scale plot in Fig. \ref{figure_GapList}(b). Similarly, the band gaps of the AF and FM states under varying doping configurations are presented in Fig. \ref{figure_GapList}(c) and (d), respectively. Based on the distinct energy and band-gap behaviors induced by different doping geometries, the doped ZGNR systems can be categorized into four typical types.

(i) First, positive $\Delta E$ values, corresponding to the red dotted regions in Fig. \ref{figure_GapList}(b), indicate that the AF state serves as the ground state for the doped ZGNRs. Such doping configurations predominantly occur under the condition of $k_{N}\approx k_{B}$. These doped ZGNRs exhibit electronic properties analogous to those of pristine ZGNRs, with non-zero band gaps for their AF states. Specifically, for configurations satisfying the positional relation $\mathbf{r}_{N}-\mathbf{r}_{B}=\pm\frac{a_{0}}{2}\hat{x}\pm\frac{\sqrt{3}a_{0}}{2}\hat{y}$ (where the carbon-carbon bond length $a_{0}=1.42$ $\AA$), the AF-state band gap can reach up to $0.2$ $eV$, while the remaining doping configurations yield a smaller band gap below $0.1$ $eV$. In contrast, the band gap of the FM state remains closed for all aforementioned configurations.

(ii) Second, negative $\Delta E$ values, marked by the blue dots in Fig. \ref{figure_GapList}(b), suggest that the FM state becomes the ground state of the doped ZGNRs. This type of doping configuration emerges in the parameter regimes where $k_{N}(k_{B})\in[3,4]$ and $1<|k_{N}-k_{B}|<10$.

(iii) Third, near-zero $\Delta E$ values, represented by the white dots in Fig. \ref{figure_GapList}(b), lead to nearly degenerate AF and FM ground states for the doped ZGNRs. Such doping configurations are sparsely distributed in the regime with $|k_{N}-k_{B}|\approx12$.

(iv) Fourth, exactly zero $\Delta E$ values, corresponding to the black dots in Fig. \ref{figure_GapList}(b), enable perfect degeneracy between AF and FM ground states. These special doped configurations are located in two typical parameter regimes: one where $k_{N}$ or $k_{B}$ equals 2 or 19, and the other where $|k_{N}-k_{B}|>13$. Notably, for ZGNRs doped with $k_{N}\in[2,4]$ and $k_{B}\in[17,19]$ (or the symmetric case of $k_{N}\in[17,19]$ and $k_{B}\in[2,4]$), both AF and FM states possess a sizable band gap up to $0.357$ $eV$, as clearly illustrated in Fig. \ref{figure_GapList}(c) and (d), respectively.

The magnetic configurations of doped ZGNRs can be visualized via the spatial distribution of spin polarization, which is defined as $\langle m_{z}\rangle\equiv\int{(\rho_{+}-\rho_{-})dydz}$, where $\rho_{+}$ and $\rho_{-}$ denote the electron densities of spin-up and spin-down states, respectively. For several representative doping configurations, the spatial distributions of $\langle m_{z}\rangle$ and the corresponding band structures are presented in the first and second rows of Fig. \ref{figure_BandList}, respectively.

\begin{figure}[tbp]
\centering
\scalebox{0.35}{\includegraphics{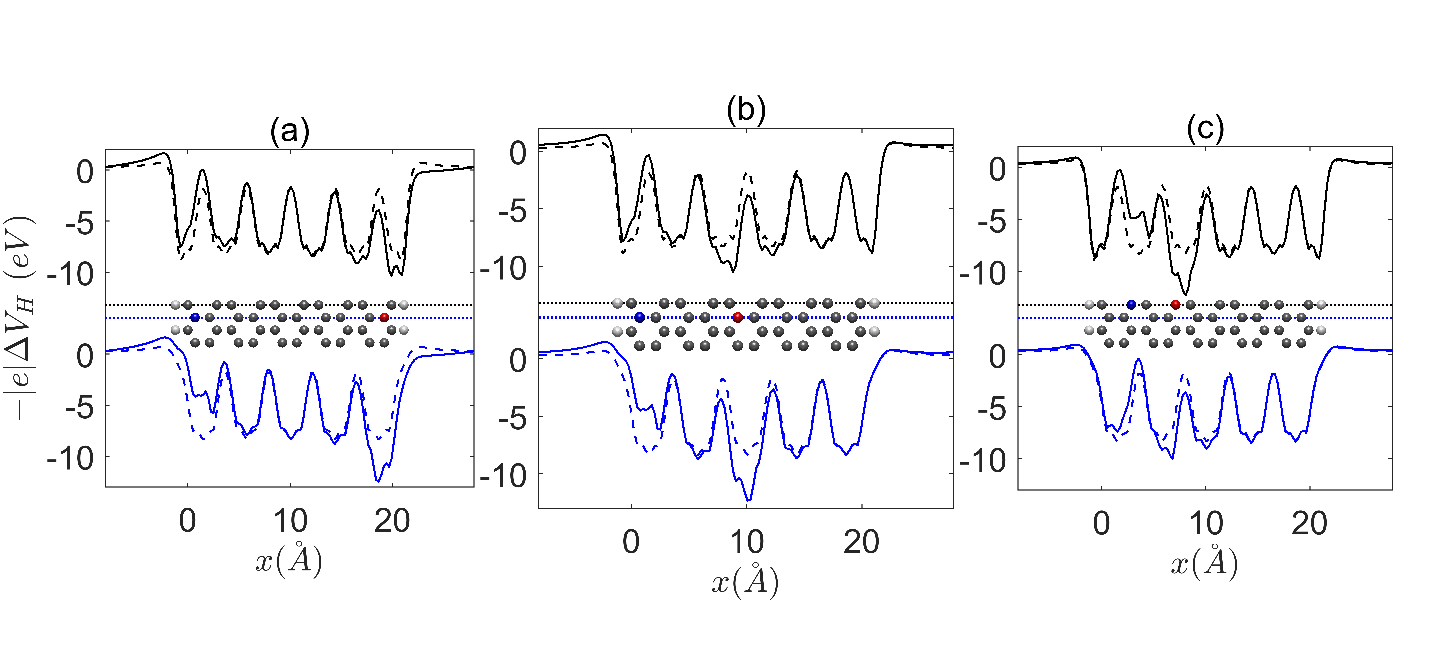}}
\caption{ Hartree difference potential along the pathes indicated by the black and blue dotted lines in the atomic structure are plotted as black and blue solid lines, respectively. The black and blue dashed lines are Hartree difference potential of pristine graphene along the same pathes. The doping configurations are $(2,19)$ in (a), $(2,10)$ in (b), and $(4,8)$ in (c).  }
\label{figure_Hpotential}
\end{figure}

For the category (iv) configurations, the AF and FM states converge to an identical nonmagnetic state originating from the demagnetization of zigzag edges. Taking the ZGNR with a typical doping configuration of $(k_{N},k_{B})=(2,19)$ as an example, both zigzag terminations are proximal to the B and N atomic arrays. Consequently, the magnetic moments at both zigzag edges are completely suppressed to zero, as illustrated in Fig. \ref{figure_BandList}(a). Such full demagnetization eliminates spin splitting in the electronic band structure, as depicted in Fig. \ref{figure_BandList}(g), while a prominent band gap is simultaneously opened. Consistent electronic and magnetic characteristics are observed for all symmetric doping cases with $k_{N}\in[2,4]$ and $k_{B}\in[17,19]$ or $k_{N}\in[17,19]$ and $k_{B}\in[2,4]$.
The Hartree difference potentials along two typical transport paths are plotted in Fig. \ref{figure_Hpotential}(a), which exhibit a steep potential gradient across the ribbon width. The substitutional N and B atoms act as n-type and p-type dopants, respectively. After charge relaxation, the local potential around N sites increases, whereas that around B sites decreases relative to pristine graphene. Quantitative analysis shows that the Hartree potential differences between the doped ZGNR and its pristine counterpart reach $\delta E_{L}=1.83$ $eV$ at the left zigzag termination and $\delta E_{R}=-2.49$ $eV$ at the right termination. Given a transverse ribbon width of $W=19.89$ $\AA$, the induced intrinsic transverse electric field is calculated as $E_{x,in}\equiv(\delta E_{L}-\delta E_{R})/W=0.2172$ $V/\AA$.
For pristine ZGNRs, spontaneous edge magnetism generates an effective exchange field at zigzag terminations, which shifts the spin-up and spin-down flat bands toward opposite sides of the Fermi level. In contrast, the intrinsic or externally applied transverse electric field modulates the local edge potential, namely the $\delta E_{L(R)}$ values. When the induced potential shift is sufficiently large to compensate the effective exchange field, both spin-polarized flat bands are displaced to the same side of the Fermi level (either entirely above or below), resulting in full depolarization of the zigzag edges \cite{maluo2021bis}.

For the doping configuration $(k_{N},k_{B})=(2,10)$, the left zigzag edge is adjacent to substitutional N dopants, which induce a substantial local potential shift $\delta E_{L}$, as demonstrated in Fig. \ref{figure_Hpotential}(b). The large potential offset fully suppresses the local edge magnetism, leading to complete demagnetization of the left zigzag termination (Fig. \ref{figure_BandList}(b)). In comparison, the right zigzag edge is far from the doped atomic sites, yielding a nearly negligible $\delta E_{R}$ and preserving robust local magnetization. The residual edge magnetism gives rise to prominent spin splitting in the corresponding band structure, as displayed in Fig. \ref{figure_BandList}(h). Owing to the asymmetric magnetic distribution with only one magnetized zigzag edge, the AF and FM states of this doped ZGNR coincide and become energetically identical.

For the ZGNR configurations categorized in types (i), (ii), and (iii), the Hartree difference potentials at both zigzag terminations remain nearly consistent with those of pristine ZGNRs, as exemplified by the representative case in Fig. \ref{figure_Hpotential}(c). Accordingly, both zigzag edges retain considerable magnetic moments, and the AF and FM states remain distinct quantum states. The modulation of $\Delta E$ by doping geometry originates from the interplay between the intrinsic internal electric field and the domain wall of the antiferromagnetic order.
For pristine ZGNRs, Lieb$^{\prime}$s theorem defines the antiferromagnetic alignment as the energetically favorable ground-state magnetic configuration, where the magnetic moment of each lattice site is oriented antiparallel to those of its nearest neighbors \cite{lieb01,lieb02,lieb03}. The left and right zigzag terminations belong to two inequivalent sublattices, naturally establishing an AF ground state with antiparallel magnetic moments at the two edges. By contrast, the FM state features a domain wall localized at the ribbon center, where parallel magnetic moments are formed between adjacent lattice sites, ultimately raising the total energy of the system \cite{maluo2020}.
Upon the introduction of paired B and N substitutional dopants, the induced internal electric field suppresses the magnetic moments at the dopant sites $k_{N}$ and $k_{B}$, as well as at the intermediate lattice sites between the two dopants, which elevates the local antiferromagnetic coupling energy. The degree of magnetic suppression differs between AF and FM states and is strongly dependent on the specific doping configuration, leading to disparate increments in local antiferromagnetic coupling energy for the two magnetic states. Consequently, the final value of $\Delta E$ is governed by the competitive interplay between the energy penalty induced by the central domain wall and the energy variation originating from dopant-induced magnetic moment suppression.

Type (i) doped ZGNRs inherit the fundamental physical characteristics of pristine ZGNRs. For type (ii) configurations, the maximum magnitude of $\Delta E$ occurs at the doping geometry of $(k_{N},k_{B})=(4,8)$, yielding an energy difference of $E_{FM}-E_{AF}=-8.95$ $meV$. The spatial distributions of spin polarization for the AF and FM states of this typical configuration are presented in Fig. \ref{figure_BandList}(c) and (d), with their corresponding electronic band structures illustrated in Fig. \ref{figure_BandList}(i) and (j), respectively. The AF state maintains a substantial $\langle m_{z}\rangle$ magnitude across the central region of the doped ZGNR, as depicted in Fig. \ref{figure_BandList}(c). In contrast, transitioning to the FM state induces prominent suppression of the central $\langle m_{z}\rangle$ amplitude due to the formation of a central domain wall (Fig. \ref{figure_BandList}(d)), which further increases the local antiferromagnetic coupling energy. Therefore, the total energy cost associated with the FM-state domain wall originates from two contributions: the elevated coupling energy caused by parallel spin alignment across the domain wall and the enhanced local antiferromagnetic coupling energy induced by suppressed central spin polarization.
Although the AF state exhibits nearly zero net magnetization due to the antiparallel arrangement of robust magnetic moments localized at the two zigzag edges, its electronic band structure still sustains significant spin splitting. Specifically, the band gap of the spin-up channel closes at the Fermi level, while the spin-down channel retains an open band gap. This distinctive band structure enables the doped ZGNR in the AF state to support pure spin-polarized transport. By comparison, the FM state possesses band structural features highly consistent with those of pristine ZGNRs.

For category (iii) configurations, the weakened magnetic moments at both zigzag edges lead to a markedly reduced $|\Delta E|$. A representative doping geometry in this category is $(k_{N},k_{B})=(5,17)$, which retains moderate edge magnetic moments while yielding an extremely small energy difference of $|\Delta E|<0.05$ $meV$. The spatial spin polarization distributions of the AF and FM states for this configuration are displayed in Fig. \ref{figure_BandList}(e) and (f), and their corresponding electronic band structures are presented in Fig. \ref{figure_BandList}(k) and (l), respectively. The intrinsic internal electric field induced by B-N doping suppresses the magnetic moments not only at the dopant sites but also across the zigzag edges and the central region of the doped ZGNR. For both AF and FM states, the $\langle m_{z}\rangle$ magnitudes in the ribbon middle region are nearly identical and far lower than the magnetic moments localized at the zigzag edges. This significant magnetic suppression greatly diminishes the domain-wall-associated energy penalty of the FM state, rendering the resultant energy cost much lower than that of category (i) and (ii) doped ZGNRs.
The physical origin of the negligible $|\Delta E|$ in category (iii) can be clarified by comparing the band structures of the $(5,17)$ and $(4,8)$ doped configurations. The edge $\langle m_{z}\rangle$ magnitudes of the $(5,17)$ configuration are approximately eight times smaller than those of the $(4,8)$ counterpart, which substantially weakens the spin splitting of its band structure, as evidenced by the subtle spin-dependent band separation in Fig. \ref{figure_BandList}(k,l) relative to the prominent splitting in Fig. \ref{figure_BandList}(i,j). For the $(4,8)$ doped ZGNR, strong interband coupling occurs between the electronic states of the two zigzag edges, leading to pronounced discrepancies in the near-Fermi-level band profiles of AF and FM states and thus producing a sizable $|\Delta E|$. In sharp contrast, the zigzag edge bands of the $(5,17)$ configuration exhibit no orbital overlap between the two edges. Transitioning from the AF to FM state merely reverses the spin polarization of the edge bands without altering the overall band morphology, which ultimately results in a near-zero $|\Delta E|$ and nearly degenerate AF/FM ground states.

\subsection{Modulation by external electric field}

\begin{figure}[tbp]
\centering
\scalebox{0.58}{\includegraphics{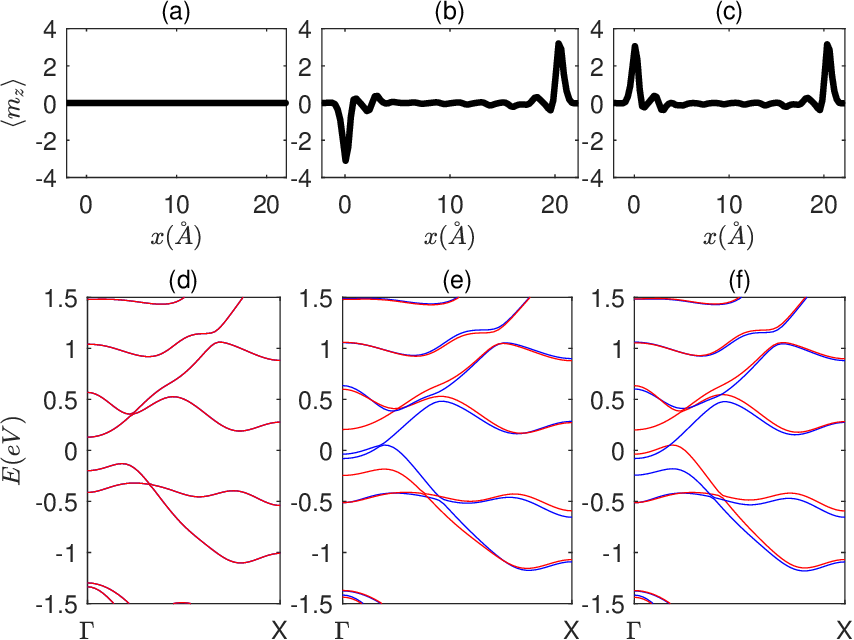}}
\caption{ For ZGNR with doping configuration $(k_{N},k_{B})=(5,17)$, in the presence of a static electric field along x direction with $E_{x}=0.94$ $V/nm$, the spatial distribution of magnetization and band structure are plotted in (a) and (d), respectively. When the direction of the static electric field is reversed to be $E_{x}=-0.94$ $V/nm$, the spatial distribution of magnetization and band structure for AF configuration are plotted in (b) and (e), respectively; those for FM configuration are plotted in (c) and (f), respectively. In the second row, the band structures of spin up and down are plotted as blue and red lines, respectively. }
\label{figure_GateBandList}
\end{figure}

The application of a transverse external electric field along the $x$-direction enables effective modulation of the magnetic configurations and electronic band structures of the doped ZGNRs. The field-dependent magnetic characteristics and band structures of the representative $(k_{N},k_{B})=(5,17)$ doped ZGNR are summarized in Fig. \ref{figure_GateBandList}. The intrinsic internal electric field of this doped system inherently weakens the zigzag edge magnetism. Accordingly, the parallel or antiparallel alignment of the external electric field with the internal electric field can respectively amplify or attenuate the total transverse electric field, thereby further suppressing or enhancing the edge magnetic moments.
When a positive external electric field of $E_{x}=0.94$ $V/nm$ is applied, the edge magnetism is completely eliminated, as illustrated in Fig. \ref{figure_GateBandList}(a). Such full magnetic suppression removes spin polarization from the electronic structure, yielding fully spin-degenerate band characteristics with no observable spin splitting (Fig. \ref{figure_GateBandList}(d)). In contrast, reversing the external electric field to $E_{x}=-0.94$ $V/nm$ strengthens the zigzag edge magnetization. The corresponding spin polarization distributions of the AF and FM states are presented in Fig. \ref{figure_GateBandList}(b) and (c), respectively, and the resultant band structures exhibit considerably enhanced spin splitting compared with the zero-field counterparts in Fig. \ref{figure_BandList}(k) and (l).
Despite the orbital overlap of edge bands under negative-bias electric fields, the interband coupling between the electronic states of the two zigzag edges remains negligible. Consequently, the energy difference $|\Delta E|$ retains a small magnitude, preserving the nearly degenerate feature of the AF and FM ground states.

\section{thermal spin and charge current}

To investigate the spin caloritronic properties of doped ZGNRs, a ballistic transport model is established for electronic conductivity calculations. In this model, both the central scattering region and the left/right electrode leads are constructed using periodic structures of doped ZGNRs along the transport $y$-direction. Under the condition of coherent ballistic transport, spin-resolved conductances for spin-up and spin-down channels can be directly derived from first-principles band structures based on the Landauer-B$\ddot{u}$ttiker formalism. Two key prerequisites are required to satisfy the coherent ballistic transport limit: the scattering region is defect-free, and its longitudinal dimension along the $y$-axis is smaller than the electron phase-coherence length.
Based on the mode-matching technique, the energy-dependent transmission probability spectrum $\mathcal{T}(E)$ is determined by counting the number of forward-propagating Bloch modes at a given energy $E$. Specifically, this quantity corresponds to the number of intersection points between the constant-energy horizontal line and the positively dispersive band branches of the electronic band structure \cite{transportLB001,transportLB002}. In the absence of spin-orbit coupling, spin-up and spin-down channels remain electronically decoupled. Accordingly, the spin-resolved transmission probabilities $\mathcal{T}_{\uparrow}(E)$ and $\mathcal{T}_{\downarrow}(E)$ are individually evaluated by identifying the intersections of the constant-energy line with the corresponding spin-polarized band dispersion curves.
The thermally driven charge current carried by spin-$\sigma$ carriers ($\sigma=\uparrow,\downarrow$) is formulated as:
\begin{equation}
I_{\sigma}=\frac{e}{h}\int{\mathcal{T}_{\sigma}(E)[f_{L}(E,\mu_{L},T_{L})-f_{R}(E,\mu_{R},T_{R})]dE}
\end{equation}
where $\mu_{L(R)}$ and $T_{L(R)}$ represent the chemical potential and temperature of the left (right) electrode lead, respectively, and $f_{L(R)}$ denotes the corresponding Fermi-Dirac distribution function for each lead. The total thermal charge current and pure spin current are further defined as:
\begin{equation}
I_{C}=I_{\uparrow}+I_{\downarrow}
\end{equation}
\begin{equation}
I_{S}=I_{\uparrow}-I_{\downarrow}
\end{equation}
respectively.

\begin{figure}[tbp]
\centering
\scalebox{0.58}{\includegraphics{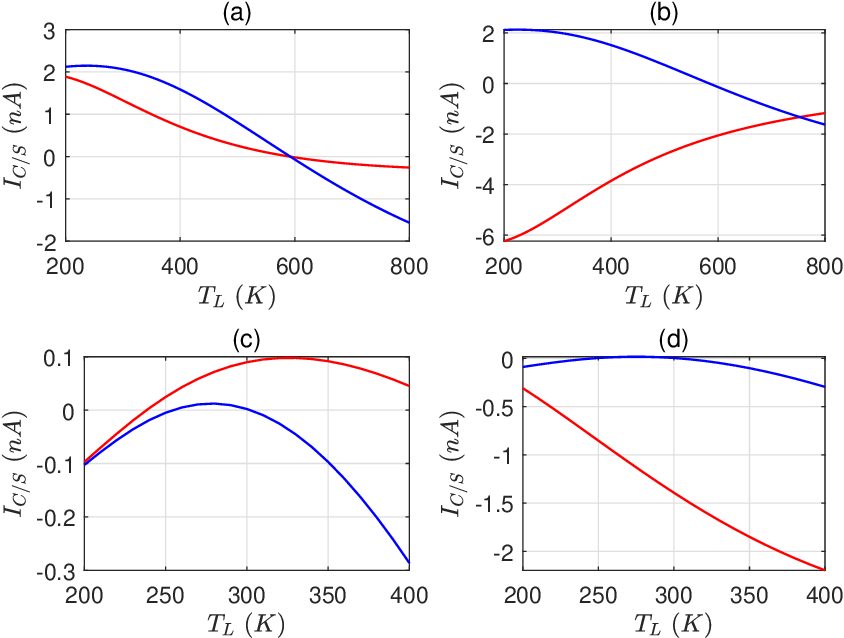}}
\caption{ Thermal charge and spin current versus $T_{L}$ with $\Delta T=1$ $K$ for ZGNR with doping configuration being $(k_{N},k_{B})=(5,17)$ are plotted as blue and red lines, respectively. The external static electric field is zero in (a,b), and $E_{x}=-0.94$ $V/nm$ in (c,d). The magnetic configuration is AF in (a,c), and FM in (b,d). }
\label{figure_ThermCurrent}
\end{figure}

For unbiased transport systems, the two electrode leads share an identical chemical potential, namely $\mu_{L}=\mu_{R}\equiv\mu$. By defining a positive temperature gradient as $\Delta T\equiv T_{R}-T_{L}>0$, the term $f_{L}-f_{R}$ becomes positive for $E<\mu$ and negative for $E>\mu$. Accordingly, the thermal current of each spin channel is collectively determined by the discrepancy in energy-resolved transmission probabilities below and above the chemical potential. Most previous studies have tuned the chemical potential near the band edge to amplify such transmission differences, thereby achieving enhanced thermal current responses. Nevertheless, the zigzag edge magnetism of doped ZGNRs exhibits strong chemical potential dependence. When the chemical potential exceeds half the maximum spin splitting energy of the edge bands, complete edge demagnetization occurs, accompanied by the disappearance of band spin splitting and full suppression of the thermal spin current. For this reason, all calculations in the present work are performed at $\mu=0$.

Figure \ref{figure_ThermCurrent} depicts the thermal charge and spin current behaviors of the representative category (iii) doped ZGNR with $(k_{N},k_{B})=(5,17)$ as functions of the left-lead temperature $T_{L}$, with a fixed temperature difference of $\Delta T=1$ K. In the absence of transverse external electric fields, the thermal charge and spin currents of the doped ZGNR under AF and FM configurations are presented in Figs. \ref{figure_ThermCurrent}(a) and (b), respectively. The numerical results reveal that the thermal charge current exhibits weak sensitivity to magnetic configurations, whereas the thermal spin current is strongly dependent on the spin-polarized magnetic states. At $T_{L}=595$ K, the thermal charge current completely vanishes, yielding a pure thermal spin current. Specifically, the pure spin current is nearly suppressed to zero for the AF state, while a considerable pure spin current of $-2$ nA is obtained for the FM state. This distinct discrepancy enables reliable on-off modulation of the pure thermal spin current via magnetic configuration switching.
When a transverse external electric field of $E_{x}=-0.94$ V/nm is applied, the aforementioned modulation capability is well maintained over a wide temperature range covering room temperature. For $T_{L}$ ranging from 250 K to 350 K, the thermal charge current is consistently suppressed below 0.01 nA. Meanwhile, the thermal spin current remains lower than 0.1 nA for the AF state but exceeds 0.8 nA for the FM state, as illustrated in Figs. \ref{figure_ThermCurrent}(c) and (d). Furthermore, such effective modulation of pure thermal spin current synergistically governed by transverse electric fields and magnetic configuration switching is generalizable to various other doped ZGNR geometries, offering a versatile strategy for designing high-performance spin caloritronic devices.

\section{conclusion}

In summary, periodic boron-nitrogen doping of ZGNRs enables effective modulation of intrinsic zigzag edge magnetism through the formation of a built-in transverse electric field. Doped ZGNRs with tailored doping geometries can be categorized into four distinct magnetic ground-state types, namely non-magnetic, antiferromagnetic, ferromagnetic, and degenerate antiferromagnetic/ferromagnetic paired states. The application of an external transverse electric field further regulates the internal electric field distribution, which modifies the edge magnetic characteristics and reshapes the spin-split band structures of the doped systems. Through rational manipulation of doping configurations and external electric field modulation, pure thermal spin current can be realized at room temperature. Notably, the resultant pure spin current possesses reversible on-off switching functionality, which can be readily toggled by transitioning between antiferromagnetic and ferromagnetic magnetic phases. The proposed B-N codoped ZGNR systems serve as robust and feasible prototypes for nanoscale spin caloritronic devices, holding great promise for the development of next-generation low-power spintronic integrated circuits.

\begin{acknowledgments}
This project is supported by  the Special Projects in Key Fields of Ordinary Universities in Guangdong Province(New Generation Information Technology, Grant No. 2023ZDZX1007), the Natural Science Foundation of Guangdong Province of China (Grant No. 2026A1515012428), and the Startup Grant at Guangdong Polytechnic Normal University (Grant No. 2021SDKYA117).
\end{acknowledgments}

\section{Appendix}

\begin{figure}[tbp]
\centering
\scalebox{0.58}{\includegraphics{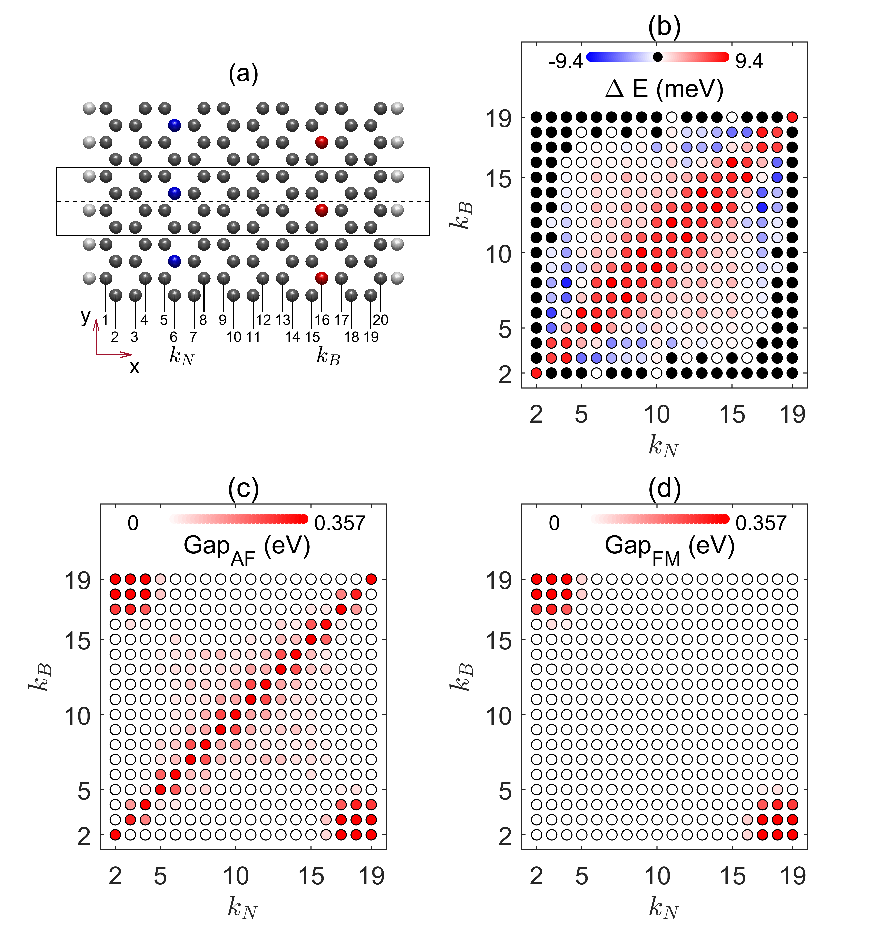}}
\caption{ The same as those in Fig. \ref{figure_GapList}, except that the boron and nitrogen dopant atoms reside in difference unit cell of the pristine ZGNR.  }
\label{figure_GapListA}
\end{figure}

Fig. \ref{figure_GapListA}(a) presents the atomic structure of doped ZGNRs where boron and nitrogen dopants are distributed across different unit cells of the pristine ribbon. In such a doping scenario, the configuration with $k_{N}=k_{B}$ is physically valid. The calculated $\Delta E$ diagram as a function of doping geometry (Fig. \ref{figure_GapListA}(b)), together with the band gaps of the corresponding AF and FM states (Fig. \ref{figure_GapListA}(c,d)), demonstrates that these doped ZGNRs can also be classified into four distinct magnetic categories. Their parameter-space distribution exhibits a highly consistent pattern with that observed for dopants confined within a single unit cell in Fig. \ref{figure_GapList}.

\clearpage

\end{document}